\pdfoutput=1
\PassOptionsToPackage{prologue,dvipsnames}{xcolor}
\documentclass[11pt]{article}
\usepackage{ACL2023}

\usepackage{times}
\usepackage{latexsym}
\usepackage{natbib} 
\usepackage{tipa}
\usepackage{graphicx}

\usepackage[T1]{fontenc}

\usepackage[utf8]{inputenc}

\usepackage{microtype}

\usepackage{inconsolata}
\usepackage{subcaption}
\usepackage{float}
\usepackage[dvipsnames]{xcolor}

\definecolor{Mild}{rgb}{1.0, 1.0, 0.718}

\title{{Probing for Phonology in Self-Supervised Speech Representations: A Case Study on Accent Perception}  }

\author{
 \textbf{Nitin Venkateswaran\textsuperscript{1}},
 \textbf{Kevin Tang\textsuperscript{2}},
 \textbf{Ratree Wayland\textsuperscript{1}}
\\
\\
 \textsuperscript{1}Department of Linguistics, University of Florida \\\texttt{\{venkateswaran.n,ratree\}@ufl.edu}\\
 \textsuperscript{2}Department of English Language and Linguistics, Heinrich Heine University Düsseldorf\\\texttt{kevin.tang@hhu.de}
}

\begin{document}
\maketitle
\begin{abstract}

Traditional models of accent perception underestimate the role of gradient variations in phonological features which listeners rely upon for their accent judgments. We investigate how pretrained representations from current self-supervised learning (SSL) models of speech encode phonological feature-level variations that influence the perception of segmental accent. We focus on three segments: the labiodental approximant [\textscriptv], the rhotic tap [\textfishhookr], and the retroflex stop [\textrtailt], which are uniformly produced in the English of native speakers of Hindi as well as other languages in the Indian sub-continent. We use the CSLU Foreign Accented English corpus \cite{l07} to extract, for these segments, phonological feature probabilities using Phonet \cite{vasquezcorrea19_interspeech} and pretrained representations from Wav2Vec2-BERT \cite{barrault2023seamless} and WavLM \cite{chen2022wavlm} along with accent judgements by native speakers of American English. Probing analyses show that accent strength is best predicted by a subset of the segment's pretrained representation features, in which perceptually salient phonological features that contrast the expected American English and realized non-native English segments are given prominent weighting. A multinomial logistic regression of pretrained representation-based segment distances from American and Indian English baselines on accent ratings reveals strong associations between the odds of accent strength and distances from the baselines, in the expected directions. These results highlight the value of self-supervised speech representations for modeling accent perception using interpretable phonological features.

\end{abstract}

\section{Introduction}
Accent perception is a complex process that is influenced not only by which speech segments a speaker produces, but also by how those segments deviate in their phonetic and phonological properties from native norms. Traditional approaches to modeling accentedness often focus on categorical segmental substitutions \cite{6497525,WIELING2012307}, or variations in segment-specific acoustic/phonetic features such as the voice onset time (VOT) of stops \cite{mccullough2013acoustic,hansen2010automatic}, the F1/F2 formants of vowels \cite{CHAN2019100919,wayland1997non} and the F2/F3 formants of liquids \cite{koffi2016acoustic,solon2015l2}.  Such models either obscure the underlying features of the segment or rely on features that do not generalize across segment types. An alternative approach is to model accentedness in terms of gradient variations in the phonological features of the segment, which reflect articulatory and acoustic properties that listeners may be more directly sensitive to \cite{hale2008phonological,METZE2007348,scharenborg2006capturing}. Phonological features are also interpretable and generalizable across segment types and languages, making them a compelling basis for cross-linguistic analysis \citep{chomsky1968sound,hayes2011introductory}.

Speech representations from self-supervised learning (SSL) models achieve state-of-the-art results on a number of speech processing benchmarks \cite{yang2021superb,shi2023ml}, including accent classification \cite{zuluaga2023commonaccent}, and are at the center of current speech-related research (for a review, see \citealt{mohamed2022self}). Recent approaches \cite{bartelds2022neural,chernyak2024perceptual} have used these representations to model accent perception using word- and utterance-level deviations between native and non-native speech. The nature of these representations is not well understood and it is unclear how such state-of-the-art results are achieved, which has resulted in a number of studies that probe SSL representations for acoustic, phonetic, speaker and language related features relevant to speech processing tasks \cite{wang2025normalization,dixit2024explaining,de2022probing,9688093}. Such studies have not paid as much attention to the extent of phonological feature information present in SSL representations that is leveraged for accent discrimination. Moreover, there is limited research modeling accentedness using segmental SSL representations, and it is unclear whether these representations can capture the effects of deviations from native segmental norms on accent strength. Given these research gaps, this study investigates the following questions:
\begin{itemize}
    \item Whether SSL representations of non-native English speech encode the gradient phonological feature variations required to model accent discrimination.
    \item Whether segment-level differences between SSL representations of native and non-native English speech predict perceived accentedness ratings.
\end{itemize}

Specifically, we focus on the investigation of segmental accent in the English of native Hindi speakers against the norms of expected segments in American English, which we use as the native norms in this study given that the accent judgements are from native speakers of American English. We select three segments: the labiodental approximant [\textscriptv], realized in place of the voiced labial fricative [v] of American English \cite{fuchs2019almost,pandey2015indian}; the rhotic tap [\textfishhookr], realized in place of the rhotic approximant [\textturnr] \cite{w15,k03,m91};  and the retroflex stop [\textrtailt], realized in place of the alveolar stop [t] \cite{m91,Kachru_1986}. These segments are selected as they are more uniformly produced in the non-native English of speakers of Hindi as well as other languages in the Indian sub-continent \cite{Wiltshire_2020,w15,fuchs2019almost,Sirsa2013TheEO,wilt,k03} and as such are more reliable indicators of accented speech across speaker productions.

We use Phonet \cite{vasquezcorrea19_interspeech} to model the selected segments as probabilistic encodings of phonological features and run probing analyses to determine the types and weights of the features present in the segments' pretrained representations that are most relevant for accent discrimination. Our results show that subsets of pretrained representation features prioritize phonological features that contrast the expected native and realized non-native English segments, only when the contrasting features are perceptually salient. We also calculate Euclidean distances from baseline American English and Indian English speech data using the pretrained representations of the selected segments, and run a regression of the distances on accent ratings. We use the term Indian English to refer to the Englishes of native speakers of languages found across the Indian sub-continent, including Hindi. An Indian English baseline is used since the available data specific to Hindi speakers is insufficient; given the wide uniformity of the selected segments' productions, an Indian English baseline may be an adequate proxy. We find the odds of strong accents increase with greater distances from American English and decrease with greater distances from Indian English. Our findings offer new insight into the ability of self-supervised speech representations to model the relationship between phonological systems and the perception of accented speech.

\section{Related Work}

Representations from self-supervised speech learning (SSL) models have been used for accent classification and identification, with features for these tasks derived from encodings of all parts of the speech utterance \cite{10887674,10773414,li2023self,deng2021improving}. While these approaches are effective, they lack explainability given that speech utterances can contain several drivers of accent perception, viz., segmental \cite{behrman2014segmental,mccullough2013acoustic,wayland1997non,munro1993productions}, suprasegmental \cite{pickering2014suprasegmental,kang2010suprasegmental,WaylandTangSengupta_AcquisitionRhythmicClass_CUP2024Accepted}, and speaker-independent factors such as lexical frequency \cite{porretta2015perceived,levi2007speaker}, and the weighting assigned by these approaches to each driver is poorly understood. In addition, the use of more abstract phonological features to analyze accents is relatively understudied (e.g., see \citealt{SANGWAN201240}). In comparison, we narrow our focus to the segmental level and leverage the rich pretrained SSL representations of the segments as features to analyse variations in accent perception, probing for phonological feature information in the representations to support more generalizable and explainable analyses. Such information has been shown to effectively capture gradient variation in both second language and clinical speech, offering interpretable dimensions of variation that may align with perceptual salience \cite{Tang2024_JASA,tang,languages8020098}. 

Prior work has modeled associations of gradient variation in speech with accentedness and intelligibility using machine learning techniques. For instance, \citet{yl11} tracked degrees of /l/-darkness using log-probabilities from forced alignments, while \citet{larty} used MFCCs to classify rhoticity using an SVM classifier. \citet{chernyak2024perceptual} project representations of the entire speech utterance from HuBERT \cite{hsu2021hubert} into a lower-dimensional space and associate distances between L1 and L2 English utterances in this space with intelligibility scores. \citet{bartelds2022neural}, which is most similar to this study, use pretrained representations from a variety of self-supervised models to calculate the average word-based distances between English utterances by native speakers of American English and Dutch, followed by the Pearson correlation coefficients between the distances and accent ratings. In comparison, we model associations between accent ratings and distances based on segment-level representations across accented speaker groups. This approach is broadly consistent with models of L2 speech production and perception, such as the Speech Learning Model (SLM/SLM-r: \citealt{f95,flege2021revised}); the SLM posits that learners of an L2 language form new L2 speech categories as a function of increasing exposure to the language, and the new categories eventually replace the learner's usage of the native L1 categories in L2 speech, resulting in the perception of a negligible/nil accent.

\section{Methods}

\subsection{Phonological features}
Speech segments can be categorised into phonological features based on their shared articulatory/phonetic features \cite{hayes2011introductory}. For instance, consonants are marked [+consonantal] as this feature involves constrictions of articulators in the vocal tract that are common across consonants; vowels and glides are [-consonantal] as they lack these constrictions. The phonological features that contrast the native and non-native [v]-[\textscriptv], [\textturnr]-[\textfishhookr], and [t]-[\textrtailt], segment pairs are of particular interest in this study, as differences in the contrastive feature values would drive changes in perceived accentness; these features are listed in Table~\ref{tab:classes}. Feature values are binary, indicating the presence or absence of the articulatory/phonetic feature. However, this binary encoding may not contain enough information to discriminate among fine-grained accent ratings given that binary values impose a categorical distinction between the native and non-native segments in the pair, which would obscure gradient feature variations driving fine-grained accent perception. We instead use Phonet (described in Section \ref{sec:probfeatureencoding}) to derive a probabilistic encoding to capture gradient variations of phonological features and their associations with the fine-grained accent ratings. 

\begin{table}[ht!]
\footnotesize
\centering
\begin{tabular}{p{0.18\columnwidth}p{0.35\columnwidth}p{0.3\columnwidth}}
\textbf{Segment Pair} & \textbf{Contrastive features}  & \textbf{Non-contrastive features}\\
\hline
\hline
[v]-[\textscriptv] &  [approximant] &[continuant]\\
 & [consonantal]&[delayed release]\\
 & [sonorant]&[labial] [voice]\\
 & &[labiodental] \\ 

\hline
[\textturnr]-[\textfishhookr] & [anterior] &[approximant] \\
 & [consonantal] &[continuant] \\
 & [tap]&[voice] [sonorant]\\
 & [distributed] &[coronal] \\

\hline
[t]-[\textrtailt] & [anterior]&[consonantal] [coronal]\\
\end{tabular}
\caption{Phonological features from \citet{hayes2011introductory} that define both segments in the native-nonnative pair.
}
\label{tab:classes}
\end{table}

\subsection{Probabilistic phonological feature encodings}\label{sec:probfeatureencoding}
Phonet \cite{vasquezcorrea19_interspeech} is a bi-directional GRU-based neural network that estimates the probabilities of phonological features associated with phonetic segments in the speech signal. The model takes MFCC-transformed speech signals and outputs a vector of phonological feature probabilities for each input frame. A single vector representation for the segment is derived by averaging the vectors across frames aligned with the segment. More details about the Phonet architecture and speech signal transformations can be found in \citet{vasquezcorrea19_interspeech}. To label the phonological features associated with each frame, a mapping between the segments and features is created for both native and non-native English phone sets, shown in Appendix~\ref{sec:appendix:phontoclassmapping} (Table~\ref{tab:mapping}). A single Phonet model is trained on the combined American and Indian English training datasets (described in Section \ref{sec:datasets}) to estimate the phonological feature probabilities for segments of both Englishes in a joint vector space. The model can be said to incorporate the acoustic properties of both Englishes in its parameter weights; given a phone segment from an accented speaker's utterance, the model can estimate whether the feature probabilities of that segment tend towards native or non-native English or contain elements of both. An 80-20 train-test split is used for training; the range of accuracy and F1 scores across the phonological features is present in in the Appendix~\ref{sec:appendix:phonetaccuracyf1} (Table~\ref{tab:scores}). The model is trained for a maximum of 30 epochs with early stopping, using the Adam optimizer \cite{Kingma2014AdamAM} with a categorical cross-entropy loss function.

\subsection{MFA pre-processing}\label{sec:mfapreprocessing}
 We use the Montreal Forced Aligner (MFA; \citealt{msmws17}) to generate phone segment-level alignments with speech signal frames. The pre-trained MFA grapheme-to-phoneme models and pronunciation dictionaries for American and Indian English are used for IPA conversion \cite{mfa_english_india_mfa_g2p_2023,  mfa_english_us_mfa_g2p_2023,mfa_english_india_mfa_dictionary_2024,mfa_english_us_mfa_dictionary_2024}. Custom acoustic models for American and Indian English are trained to avoid potentially noisy output from the existing pre-trained model \cite{mfa_english_mfa_acoustic_2024}, given that this model is trained on a variety of world Englishes. 

\subsection{Datasets}\label{sec:datasets}
Three separate datasets are used in this study. Two large speech corpora of American English and Indian English serve as baselines, and the CSLU FAE Release 1.2 \cite{l07} provides speech samples with accent ratings for analysis.

The baseline datasets are sourced from the Mozilla Common Voice Speech Corpus \cite{ardila-etal-2020-common}, the Librispeech-100 corpus \cite{7178964}, the L2-ARCTIC non-native English speech corpus \cite{zsslclg18}, and the Indic Text-To-Speech (TTS) corpus \cite{btnm16}, and are used to train the Phonet model and provide references for the distance-based calculations described in Section \ref{sec:probinganalyses}. A total of 405 hours of combined American and Indian English data are used with a balanced split between the two varieties.

\begin{figure}[ht!]
\centering
\includegraphics[width=0.4\textwidth]{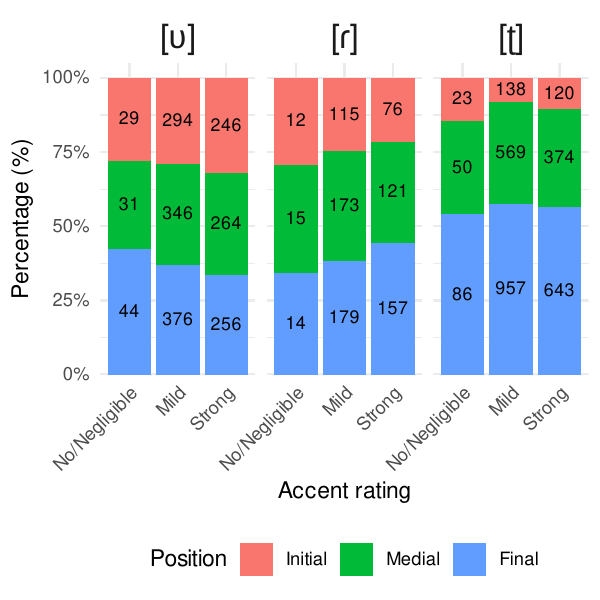}
\caption{Distribution of target segments in the CSLU FAE subset of Hindi speakers by word position and accent rating. 
   \label{tab:segmentdists}}
\end{figure}

For analysis, we use the CSLU FAE Release 1.2 dataset \cite{l07} consisting of continuous speech in English by native speakers of 22 languages along with accent annotations for each speech utterance. We focus on the subset of 349 utterances, each produced by a different native Hindi speaker.. The utterances are telephone-quality, with speakers asked to speak about themselves in English for 20 seconds. Each utterance was rated for accentedness by three native speakers of American English on a 4-point scale: \textit{1-negligible/no accent}, \textit{2-mild accent}, \textit{3-strong accent} and \textit{4-very strong accent}. To establish a conservative estimate of perceived accent strength, we adopt the minimum rating across the three raters as the final score for each recording. Due to the low resulting frequency (<1\%) of the \textit{very strong} rating, we merge that category with the \textit{strong} rating resulting in three final levels: \textit{no/negligible}, \textit{mild}, and \textit{strong}. Figure~\ref{tab:segmentdists} summarizes the distribution of the target segments in the CSLU FAE subset of Hindi speakers by accent rating and word position.

\subsection{Pretrained SSL representations}
We select the Wav2Vec2-BERT \cite{barrault2023seamless} and WavLM \cite{chen2022wavlm} architectures and extract pretrained representations from the 24 Conformer \cite{conformer} and Transformer layers of the \texttt{w2v-bert-2.0} conformer encoder and \texttt{wavlm-large} model for our experiments. 
These models have been shown to outperform similar models including Wav2Vec 2.0 \cite{baevski2020wav2vec} and HuBERT \cite{hsu2021hubert} on a variety of speech processing tasks \cite{chung2021w2v,hsu2021hubert}. We use two models here to check for the robustness of results, given that the two models differ in certain key aspects, such as in model architectures with Wav2Vec2-BERT using Conformer encoders versus Transformer encoders in WavLM, and in the training objectives with a joint contrastive and MLM-based objective used for Wav2Vec2-BERT versus masked speech prediction and denoising objectives used for WavLM \cite{chung2021w2v,chen2022wavlm}. A single segmental representation is derived by averaging the representations across all 20ms frames aligned with the segment. We use HuggingFace's \texttt{transformers} library  \cite{wolf2019huggingface} to extract the representations using the phone level alignments described in Section \ref{sec:mfapreprocessing}.

\subsection{Probing analyses}\label{sec:probinganalyses}
We use two independent accent classification models (probes) to evaluate each Conformer layer in Wav2Vec2-BERT and Transformer layer in WavLM. Separate probing classifiers are run for each segment and representation type (Wav2Vec2-BERT or WavLM). The probing classifiers take the features of the segment's pretrained representation as input to determine which layer best distinguishes different accent ratings on a test set. We use the weighted-F1 score given the class imbalance seen in Figure~\ref{tab:segmentdists}. A logistic regression and linear support vector classifier probe are used, with an L1 regularization term added to the models' objective functions to shrink the $\beta$-coefficients of non-relevant features to zero \cite{lasso}, thereby selecting only the subset of features from the segment's pretrained representation that is most relevant for accent classification. Two different probes are used to evaluate the robustness of results. The probing classifiers' scores from the best layer are compared to a baseline classification model that uses the segment's MFCC features, derived from the processing pipeline of the \texttt{w2v-bert-2.0} encoder. We use the \texttt{glmnet} package \cite{glmnet} in \texttt{R} for the logistic regression classifier and \texttt{scikit-learn} \cite{scikit-learn} for the linear support vector classifier, and run grid searches to determine the optimal regularization parameters.

To investigate whether the probes make use of phonological feature information present in the segment's pretrained representation for accent discrimination, the accent-relevant subset of features selected from the representation by the probing classifier is correlated with the segment's phonological feature probabilities independently derived by Phonet, using the \texttt{svcca} toolkit \cite{raghu2017svcca}. \texttt{svcca} calculates canonical correlations between sets of representations from deep neural networks in an efficient way that is invariant to affine transform. We calculate scores for each phonological feature by accent, representation, and probe type using the accent-relevant subsets from the best performing layer. We also calculate as an accent-agnostic baseline the canonical correlations for each phonological feature using the segment's full set of pretrained representation features. A softmax normalization is applied to convert the correlations into weights across phonological features, and relative correlation weights are calculated by taking, for each phonological feature, the ratio between the weight calculated using the accent-relevant subset and the weight from the accent-agnostic baseline. A ratio that is greater than one for a phonological feature indicates that the accent-relevant subset contains more information about that feature relative to the accent-agnostic baseline, thereby indicating that feature's relevance for accent discrimination; ratios smaller than one indicate that the feature is not relevant.

 To investigate whether the pretrained representations contain segment level differences relevant for accent classification, we also calculate the average Euclidean distance between the target segments in the CSLU FAE dataset and the native and nonnative segments from the baseline American and Indian English datasets, and run a multinomial logistic regression of the distances on the accent ratings taking the \textit{no/negligible} rating as the reference level. Interactions of distance with word position are also investigated, given that variations in the categorization of a speech segment can be driven by the position of the segment in the word sequence \cite{DMITRIEVA2019128}. All the representation features from the best layer are used in the distance calculations. The odds of a \textit{mild} and \textit{strong} accent over the \textit{no/negligible} one are expected to increase (decrease) with increasing distance from the American English (Indian English) baseline.

\section{Results}

\subsection{Layer-wise accent classification probing} \label{sec:results:layerwiseprobing}
Figure~\ref{fig:probe_logreg_r} in Appendix~\ref{sec:appendix:layerwisescores} shows the layer-wise weighted-F1 accent classification scores by representation and probe type for the target segments. The middle layers show the strongest associations with accent ratings, which aligns with existing research showing stronger accent-specific phoneme representations \cite{yang2023can} and stronger correlations between accent ratings and representation-based distances calculated using the middle layers \cite{bartelds2022neural}. Table~\ref{tab:best_layer_F_scores} shows the scores based on the pretrained representation and MFCC features of segments from the best layer. The pretrained representations outperform the MFCC baselines as expected. The Wav2Vec2-BERT representations show better accent discrimination than the WavLM ones across segment and probe types, which aligns with existing research showing that Wav2Vec2-type models outperform others in accent identification tasks \cite{10773414}.

\begin{table}[H]
\footnotesize
\centering
\begin{tabular}{cc|cc}
\textbf{Segment} & \textbf{Feature Type} & \textbf{LR} &    \textbf{SVM}  \\
\hline
\hline
[\textscriptv] & Baseline: MFCC & 41.27 & 43.9  \\ 
 &  Wav2Vec2-BERT  & \textbf{67.5} & \textbf{68.28}  \\ 
 &  WavLM  & 64.83 & 65.42   \\

\hline
[\textfishhookr]  & Baseline: MFCC & 48.5 & 46.97    \\ 
 &  Wav2Vec2-BERT  & 65.52 & \textbf{68.43}  \\ 
 &  WavLM  & \textbf{66.34} &  65.9  \\
 
\hline
[\textrtailt] &  Baseline: MFCC & 40.52 & 47.53 \\ 
 &  Wav2Vec2-BERT  & \textbf{73.1} & \textbf{74.1}  \\ 
 &  WavLM  & 69.27 & 69.28   \\

\hline

\end{tabular}
\caption{Weighted-F1 accent classification scores by feature type (MFCC, Wav2Vec2-BERT and WavLM) and probe type (LR=Logistic Regression; SVM=Linear Support Vector Machine) using the target segments' representations from the best layer.}
\label{tab:best_layer_F_scores}
\end{table}

\begin{table}[ht!]
\begin{subtable}[h]{0.45\textwidth}
\footnotesize
\centering
\begin{tabular}{lll|r|r}
\textbf{Seg.} & \textbf{Accent} & \textbf{Effect}&  \textbf{$\beta$} &  \textbf{\textit{p}-val}  \\
\hline
\hline
[\textscriptv] & Mild & AE Dist. & -1.0 & .027 \\ 
  & & IE Dist. & 0.5 & .249\\
 & & Final Pos. & -0.1 & .856\\
 &  & AE*Medial & 1.7 & .003 \\ 
 &  & AE*Final & 1.8 & .017 \\
 &  & IE*Medial & -1.4 & .011 \\
 & Strong & AE Dist. & -1.0 & .022 \\
 & & IE Dist. & 0.5 & .174\\
 &  & Medial Pos. & -0.1 & .706 \\ 
 &  & Final Pos. & -0.1 & .724 \\ 
 &  & AE*Medial & 1.7 & .004 \\
 &  & AE*Final & 2.6 & <.001 \\
 &  & IE*Medial & -1.4 & .011 \\
 &  & IE*Final & -2.1 & .004 \\
\hline
[\textfishhookr] & Mild & AE Dist. & 0.8 & <.001  \\ 
 &  & IE Dist. & -1.0 & <.001  \\
 & Strong  &  AE Dist. & 1.3 & <.001  \\
& & IE Dist. & -1.4  & <.001  \\
& & Final Pos. & 0.9  & .041  \\

\hline
[\textrtailt] & Mild & AE Dist. & 2.0 & .004 \\
 &  & IE Dist. & -1.9 & .011 \\
 & Strong & AE Dist. & 2.8 & <.001 \\
 &  & IE Dist. & -2.8 & <.001 \\
 & & Final Pos. & -0.1 & .785 \\
 & & AE*Final & 3.7 & .009\\
 & & IE*Final & -4.0 & .004\\
 
\hline

\end{tabular}
\caption{Wav2Vec2-BERT}
\label{tab:logregscoresw2v2}
\end{subtable}\vspace{0.03\textwidth}%

\begin{subtable}[h]{0.45\textwidth}
\footnotesize
\centering
\begin{tabular}{lll|r|r}
\textbf{Seg.} & \textbf{Accent} & \textbf{Effect}&  \textbf{$\beta$} &  \textbf{\textit{p}-val}  \\
\hline
\hline
[\textscriptv] & Mild & IE Dist. & -0.8 & .033 \\ 
 & Strong & AE Dist. & 0.3 & .408 \\
 &  & IE Dist. & -0.8 & .039 \\ 
 &  & Final Pos. & 1.0 & .012 \\ 
 &  & AE*Final & 2.4 & <.001 \\
 &  & IE*Final & -2.1 & .002 \\
\hline
[\textfishhookr] & Mild & AE Dist. & 2.6 & <.001  \\ 
 &  & IE Dist. & -2.2 & <.001  \\
 &  & Medial Pos. & 0.9 & .045  \\
 &  & Final Pos.& 1.9 & .002  \\
& Strong  &  AE Dist. & 3.3 & <.001  \\
& & IE Dist. & -2.8 & <.001  \\
& & Medial Pos. & 1.1 & .023  \\
& & Final Pos. & 2.4 & <.001  \\

\hline
[\textrtailt] & Mild & IE Dist. & -1.1 & .005 \\
 &  & Final Pos. & 0.7 & .013 \\
 & Strong & AE Dist. & 1.8 & <.001 \\
 &  & IE Dist. & -2.1 & <.001 \\
 
\hline

\end{tabular}
\caption{WavLM}
\label{tab:logregscoreswavlm}

\end{subtable}
\caption{Log-odds coefficients ($\beta$) from multinomial logistic regressions of distances on accent ratings, taking the \textit{no/negligible} accent as reference level. Significant main effects are reported, and main and interaction effects are reported if the interaction effect is significant(\textit{p}< .05). Positive log-odds coefficients suggest increased odds of the accent rating over the \textit{no/negligible} accent per unit increase in the regressor. Negative coefficients suggest decreased odds. (AE=American English; IE=Indian English) }
\label{tab:logregscores}
\end{table}

\begin{figure}[h!]
\includegraphics[width=0.48\textwidth]{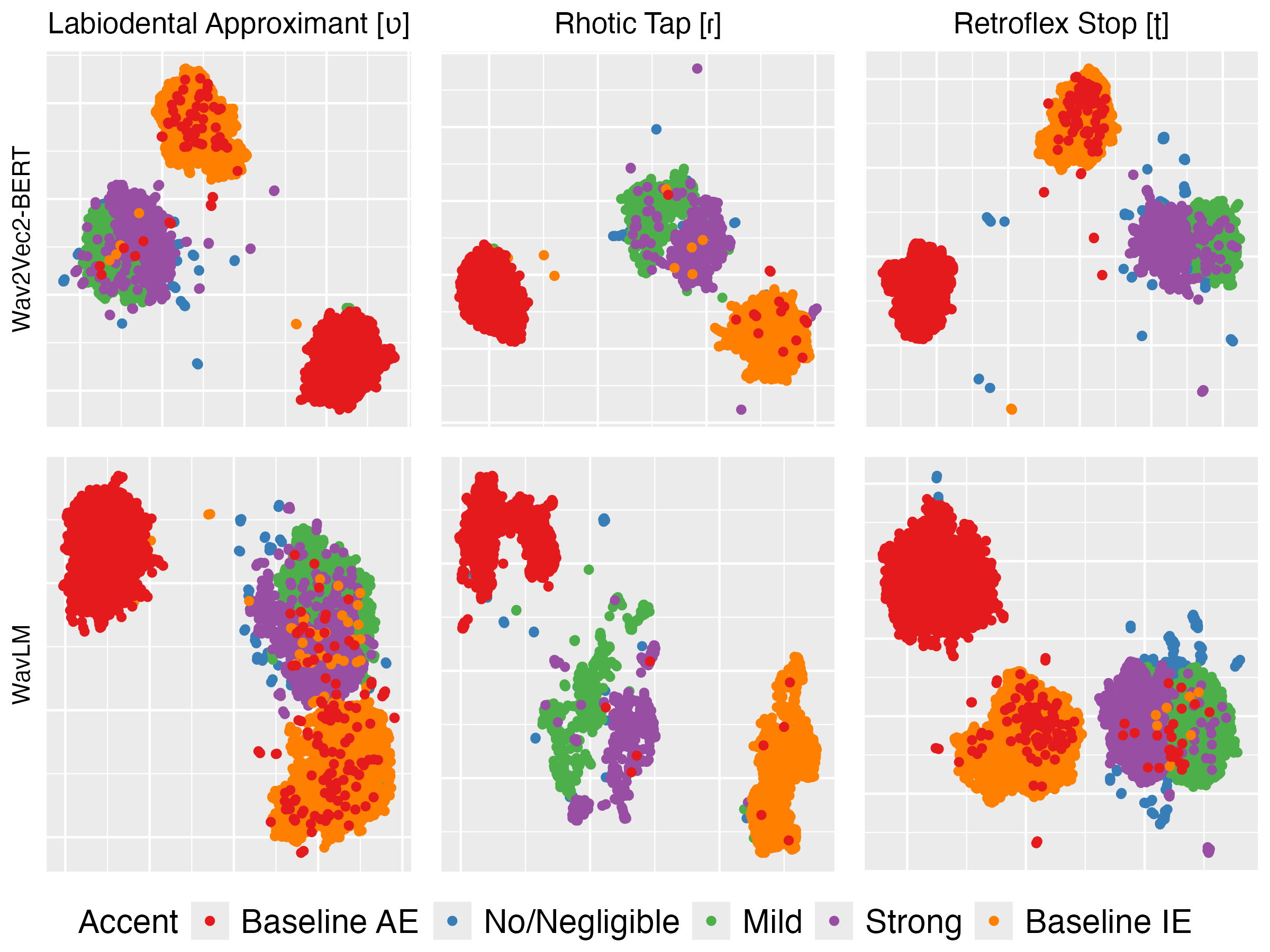}
\caption{Scatterplots of UMAP embeddings of target segments from the CSLU FAE corpus with different accent ratings (\textcolor{Blue}{\textbf{No/Negligible}}, \textcolor{Green}{\textbf{Mild}}, \textcolor{Purple}{\textbf{Strong}}) alongside embeddings of samples of baseline \textcolor{red}{\textbf{American}} and \textcolor{Orange}{\textbf{Indian}} English segments. The embeddings are derived from supervised dimension reduction of the pretrained Wav2Vec2-BERT/WavLM representations.
   \label{fig:clusters_umap}}
\end{figure}

\subsection{Distance-based associations with accent ratings} \label{sec:results:distaccentrating}

Figure~\ref{fig:clusters_umap} shows the clusters of the target segments in the CSLU FAE data by accent rating alongside the native and nonnative segment clusters from the baseline American and Indian English data. We use UMAP \cite{mcinnes2018umap} to reduce the dimensions of the pretrained segment representations extracted from the best layers, and create scatterplots of the resulting 2-dimensional embeddings. The scatterplot shows that the segments marked with the \textit{strong} accent are closer to the baseline Indian English clusters relative to segments marked with the \textit{mild} accent. Tables~\ref{tab:logregscoresw2v2} and~\ref{tab:logregscoreswavlm} show the log odds-ratios ($\beta$-coefficients) for significant main and interaction effects (\textit{p}<.05) of the multinomial logistic regression models. The results are similar across segment and representation types. For the rhotic tap [\textfishhookr] and retroflex stop [\textrtailt] segments, the odds of \textit{strong} and \textit{mild} accents increases (decreases) with increasing distance from the American English (Indian English) baseline, and the changes in odds are uniform across all word positions for the rhotic tap [\textfishhookr]; the Wav2Vec2-BERT retroflex stop [\textrtailt] shows interaction effects word-finally with distances from both American and Indian English, and the sum of $\beta$-coefficients across main and interaction effects indicates more prominent word-final effects of distance on the \textit{strong} accent rating. For the labiodental approximant [\textscriptv], both Wav2Vec2-BERT and WavLM representations show stronger word-final interaction effects with both distances on the \textit{strong} accent relative to other word positions. The Wav2Vec2-BERT representations also show word-medial interaction effects with both distances on both \textit{mild} and \textit{strong} accents, though these are weaker than the word-final interaction effects.

\subsection{Canonical correlations of phonological features with pretrained representations} \label{sec:results:cancorr}

\begin{figure*}
\includegraphics[scale=0.30]{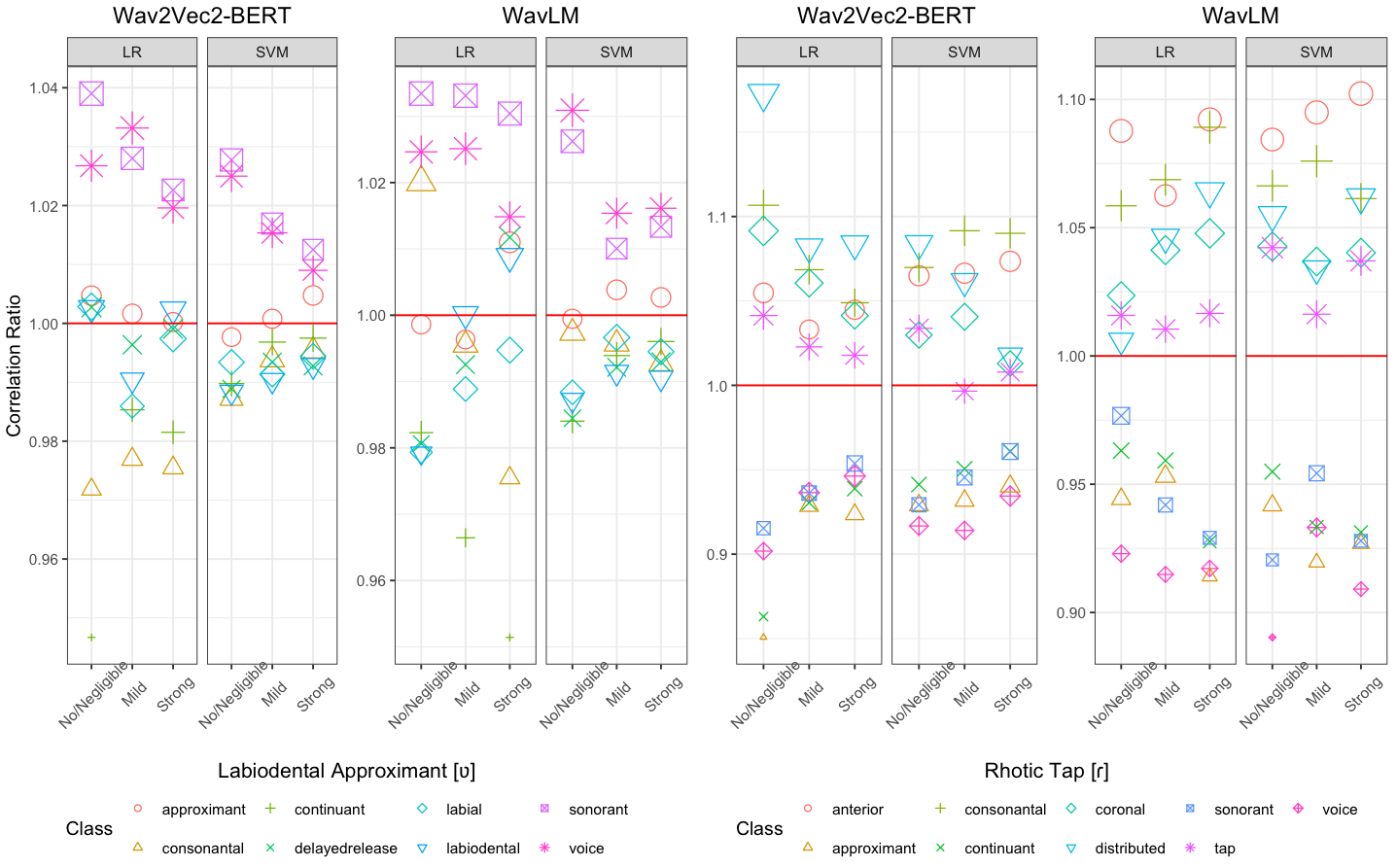}
\caption{ Canonical correlation-based relative weights of phonological features of the labiodental approximant [\textscriptv] and rhotic tap [\textfishhookr] segments, derived from accent-relevant subsets of the segment's pretrained representation. The weights reflect the prominence of the phonological feature for accent classification and are shown by accent, representation, and probe type. The red line indicates an accent-agnostic weight baseline derived from correlations with the full representation.  Phonological features with relative weights above the red baseline are given higher weighting for accent discrimination (LR=Logistic Regression; SVM=Linear Support Vector Machine).
   \label{fig:ccarv}}
\end{figure*}

\begin{figure}[H]
\includegraphics[scale=0.225]{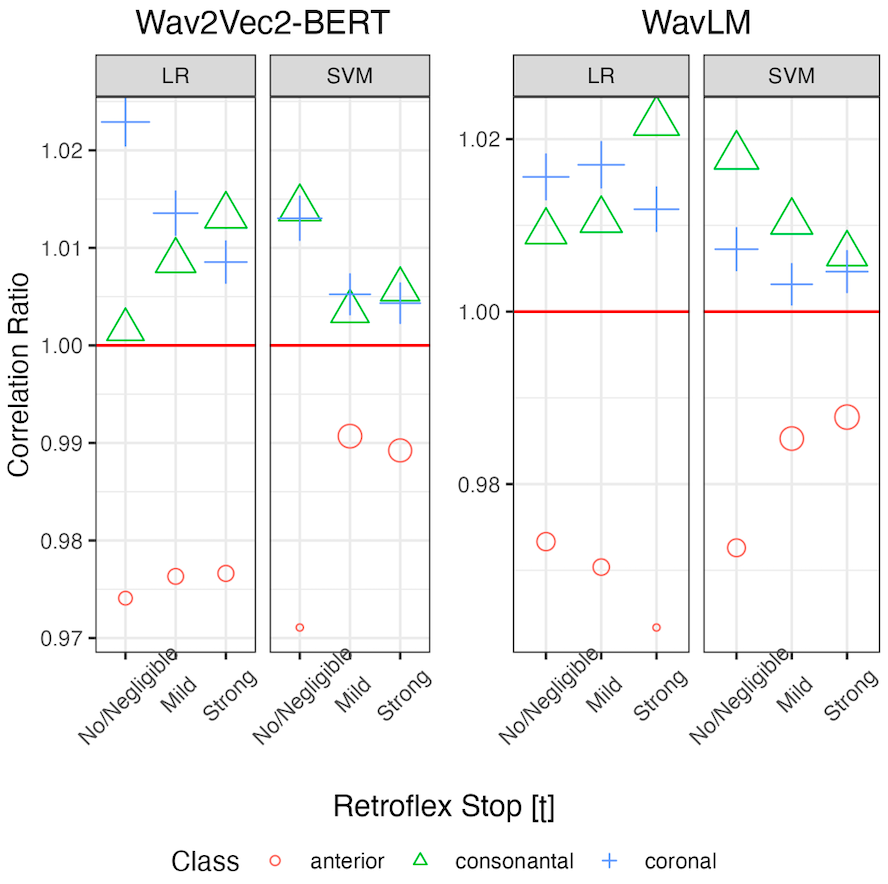}
\caption{Canonical correlation-based relative weights of phonological features of the retroflex stop [\textrtailt]  segment, derived from accent-relevant subsets of the segment's pretrained representation. The weights reflect the prominence of the phonological feature for accent classification and are shown by accent, representation, and probe type. The red line indicates an accent-agnostic weight baseline derived from correlations with the full representation.  The [anterior] feature, with relative weights below the red baseline, is not relevant for accent discrimination (LR=Logistic Regression; SVM=Linear Support Vector Machine).
   \label{fig:ccat}}
\end{figure}

Figures~\ref{fig:ccarv} and \ref{fig:ccat} show the canonical correlation-based relative weights, discussed in Section \ref{sec:probinganalyses}, of the phonological features by accent, representation, segment, and probe type. The correlation values are listed in Appendix \ref{sec:appendix:rawcancorrelationscores} (Tables~\ref{tab:ccatap}-\ref{tab:ccalabiodental}). The red baselines in the figures indicate the weights of phonological features derived from the accent-agnostic baselines; these weights have been relativized to one.  The results for the labiodental approximant [\textscriptv] show that the [sonorant] and [approximant] features contrasting the [v]-[\textscriptv] segments, as well as the [voice] feature are given larger relative weights (>1) than the other features, indicating their importance for accent discrimination. Results for the rhotic tap [\textfishhookr] show that the [anterior], [consonantal], [distributed], and [tap] features contrasting the [\textturnr]-[\textfishhookr] pair, as well as the [coronal] feature are given larger relative weights . However, looking at the retroflex stop [\textrtailt], the [anterior] feature that distinguishes the [t]-[\textrtailt] contrast does not rank as high as the other [consonantal] and [coronal] features; the [consonantal] feature here has the largest weighting, closely followed by [coronal]. 

\section{Discussion}
Our findings show that a subset of pretrained SSL representations is highly predictive of accentedness and contains relatively more information about perceptually salient phonological features that contrast the expected native and realized non-native English segments. The importance of perceptual salience is demonstrated by the [anterior] feature, which distinguishes the native alveolar stop [t] from the non-native retroflex stop [\textrtailt] but did not show the expected prominent weighting in the pretrained representations. This likely reflects American English listeners’ low perceptual sensitivity to retroflexion, a pattern well-documented in cross-language speech perception \cite{10.1121/1.2161427,polka1991cross,werker1984phonemic}. The significant effects, particularly in word-final position, of distances from the American English [t] and Indian English [\textrtailt] baselines on accent ratings further suggests that listeners are sensitive to features of the retroflex stop other than [anterior]. Looking at the rhotic tap [\textfishhookr], the high weighting assigned to the [coronal] feature suggests co-articulatory effects with the contrastive [tap] feature as well as articulatory differences in the position of the tongue tip/blade between the realizations of the rhotic approximant and tap segments. The high weighting of the [voice] feature suggests voicing differences between the fricative [v] and approximant [\textscriptv] segments that, along with the contrastive [sonorant] and [approximant] features, are salient to the perception of accent. These findings suggests that pretrained representations implicitly encode linguistically meaningful articulatory-acoustic structure for modeling accent perception.

The log-odds coefficients of the multinomial logistic regression model listed in Table~\ref{tab:logregscores} show increasing odds of a \textit{mild} and \textit{strong} segmental accent with increasing distances from American English baselines, and decreasing odds with increasing distances from Indian English baselines. These results, validated using pretrained representations in this study, align with the Perceptual Distance Hypothesis \cite{GOSLIN201292}, which states that highly proficient non-native speech that is perceptually similar to native speech is easier to process than perceptually dissimilar i.e. less proficient non-native speech, with the perceptual scale derived from the acoustic distances between native and non-native speech (for e.g. see \citealt{clarke2004rapid}). The finding that the odds of \textit{mild} and \textit{strong} accents vary depending on the position of the segment within the word is also consistent with the Speech Learning Model (SLM), which states that the use of L1 categories in L2 speech occurs at the level of position-sensitive allophones.

\section{Conclusion and Future Directions}

This study demonstrates that segmental accent perception is grounded in phonological feature-level variation that can be modeled in spontaneous non-native speech using representations from self-supervised speech learning models, which represent the current state-of-the-art in speech processing. Future work may expand these findings to prosodic dimensions or to other native–nonnative speech pairs.

\section*{Limitations}

We acknowledge the limitations in our current study that could be addressed in future research. The degree of accentedness of the Indian English baseline dataset is not available, therefore it is possible that Indian English speakers in the data might overlap with American English speakers. Figure \ref{fig:clusters_umap} shows that a small portion of American segments cluster with Indian English segments, but not the other way round. While the use of British English as the native norm might be more appropriate than American English given the colonial history of the Indian sub-continent, the use of American English was justified because the accent ratings were made by native American English speakers, and because the native Hindi speakers in the CSLU FAE corpus were resident in America at the time of recording and most had been exposed to the dominant American English environment for a number of years. Accent ratings are also affected  by other non-target segments as well as suprasegmental information such as rhythm which we did not control for.

\section*{Ethics Statement}

The models we use vary in size. We are aware of the environment impact training and inference have. Not all data used in the study are from publicly accessible datasets. The Indic-TTS corpus licensing terms allow for copies and derivatives of the data to be made, with ownership by the Licensee. The CSLU FAE Release 1.2 corpus is available via the LDC (Linguistic Data Consortium) portal and is made available for non-commercial research purposes. The CommonVoice dataset is available under the Mozilla Public License 2.0, and the other datasets under CC BY-NC 4.0. To promote transparency and reproducibility, all data, except for the Indic-TTS and CSLU FAE datasets, and code used in this study are publicly available. The involved university does not require IRB approval for this kind of study, which uses publicly available data without involving human participants. We do not see any other concrete risks concerning dual use of our research results. Of course, in the long run, any research results on AI methods could potentially be used in contexts of harmful and unsafe applications of AI. But this danger is rather low in our concrete case.

\bibliographystyle{acl_natbib}
\bibliography{custom,anthology}

\appendix

\section{Appendix}
\label{sec:appendix}

\subsection{Layer-wise segmental accent classification scores}\label{sec:appendix:layerwisescores}
\begin{figure*}
\includegraphics[scale=0.315]{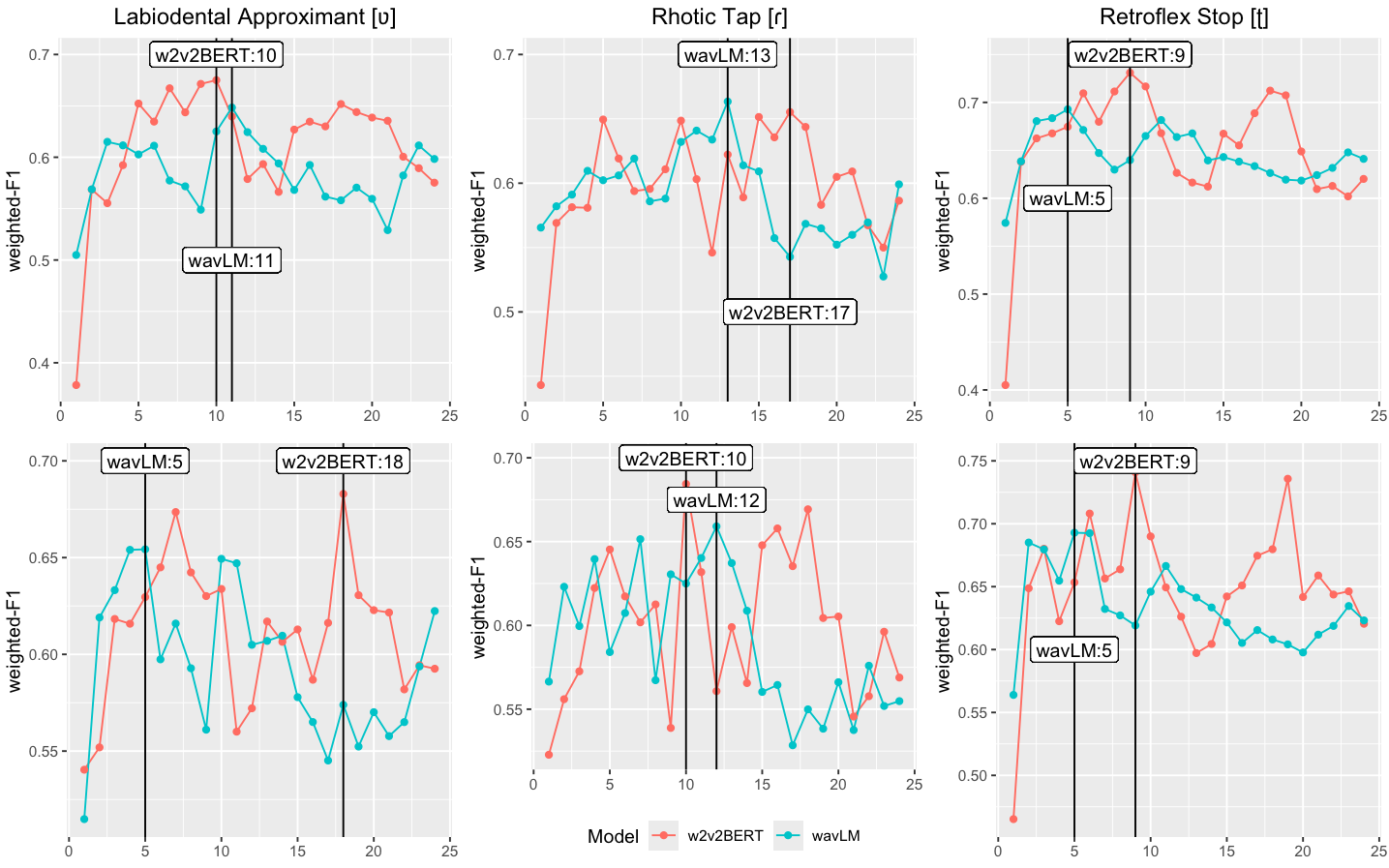}
\caption{Layer-wise weighted-F1 accent classification scores from logistic regression (top) and support vector classifier (bottom) probes that take as input the pretrained representations of the target segments extracted from Wav2Vec2-BERT (orange) and WavLM (blue). The layers with the best scores are highlighted. 
   \label{fig:probe_logreg_r}}
\end{figure*}

Figure~\ref{fig:probe_logreg_r} shows the layer-wise accent classification scores for the target segments in this study by pretrained representation and probe type.

\subsection{Phonet accuracy and F1 scores}\label{sec:appendix:phonetaccuracyf1}
Table~\ref{tab:scores} shows the Phonet model's accuracy and F1 classification scores for each phonological feature.

\begin{table}[htbp]
\centering

\begin{tabular}{l|l|l}
\textbf{Phonological Feature} & \textbf{Accuracy} & \textbf{F1 score} \\
\hline
\hline
syllabic & 91.07 & 91.23 \\
consonantal & 91.55 & 91.59 \\
long & 86.69 & 88.8 \\
sonorant & 93.68 & 93.68 \\
continuant & 92.50 & 92.50 \\
delayed release & 91.98 & 92.57 \\
approximant & 92.86 & 92.9 \\
tap & 97.31 & 98.33 \\
nasal & 91.83 & 92.98 \\
voice & 93.2 & 93.2 \\
spread glottis & 95.66 & 96.81 \\
labial & 87.65 & 88.8 \\
round & 90.4 & 92.42 \\
labiodental & 92.49 & 94.31 \\
coronal & 88.65 & 89.02 \\
anterior & 88.08 & 88.79 \\
distributed & 87.56 & 90.31  \\
strident & 95.11 & 95.52 \\
lateral & 92.9 & 94.8 \\
dorsal & 90.97 & 91.01 \\
high & 87.56 & 88.61 \\
low & 91.37 & 92.41 \\
front & 90.26 & 90.99 \\
back & 90.33 & 92.01 \\
tense & 86.84 & 90.98 \\
constr glottis & 99.99 & 99.99 \\

\hline
\end{tabular}
\caption{Accuracy and F1 scores for phonological feature classification by Phonet.}
\label{tab:scores}
\end{table}

\subsection{Raw canonical correlation scores}\label{sec:appendix:rawcancorrelationscores}
Tables~\ref{tab:ccatap}, \ref{tab:ccaretro}, and \ref{tab:ccalabiodental} show the raw canonical correlation scores between the subset of the target segments' pretrained representations selected by the accent classification probes and the phonological feature probabilities independently derived by Phonet; scores are listed by accent, pretrained representation, and probe type.

\begin{table}[htpb]
\centering
\resizebox{\columnwidth}{!}{%
\begin{tabular}{cc|cc|cc}
\multicolumn{2}{c|}{} &\multicolumn{2}{c|}{\textbf{Wav2Vec2-BERT}} & \multicolumn{2}{c}{\textbf{WavLM}} \\
\hline
 \textbf{Accent} & \textbf{Feature} & \textbf{LR} &    \textbf{SVM} & \textbf{LR} &    \textbf{SVM}  \\

\hline
 No/Negligible  & anterior & \textbf{0.504} & \textbf{0.642} &\textbf{0.436} & \textbf{0.649} \\ 
   & approximant  & 0.289 & 0.505 & 0.291 &0.504 \\ 
   & consonantal  & \textbf{0.557} & \textbf{0.651} & \textbf{0.414}& \textbf{0.638}  \\
   & continuant  & 0.30 & 0.515 & 0.311& 0.518 \\
   & coronal  & \textbf{0.536} & 0.606 & \textbf{0.376}& \textbf{0.611} \\
   & distributed  & \textbf{0.608} & \textbf{0.658}  &0.359 & \textbf{0.623}\\
   & sonorant  & 0.355 & 0.498 & 0.325& 0.482 \\
   & tap  & 0.492 & \textbf{0.613} & \textbf{0.367}& 0.608 \\
   & voice  & 0.338 & 0.483  & 0.263& 0.443\\
 \hline
 Mild  & anterior & \textbf{0.70} & \textbf{0.568} &\textbf{0.66} & \textbf{0.595}\\ 
   & approximant  & 0.593 & 0.433& 0.547&  0.416\\ 
   & consonantal  & \textbf{0.738} & \textbf{0.595} & \textbf{0.671}& \textbf{0.583} \\
   & continuant  & 0.591 & 0.449 & 0.554&0.432  \\
   & coronal  & \textbf{0.724} & \textbf{0.541} & \textbf{0.641}& \textbf{0.541} \\
   & distributed  & \textbf{0.744} & \textbf{0.561}  &\textbf{0.645} & \textbf{0.539}\\
   & sonorant  & 0.593 & 0.44  & 0.536&0.454 \\
   & tap  & 0.69 & 0.501 & 0.609& 0.519\\
   & voice  & 0.592 & 0.405 & 0.502&  0.426\\
 \hline
   Strong  & anterior & \textbf{0.71} & \textbf{0.61} &\textbf{0.705} &\textbf{0.623} \\ 
   & approximant  & 0.586 &  0.478 & 0.523&0.446 \\ 
   & consonantal  & \textbf{0.718} & \textbf{0.63} & \textbf{0.708}&  \textbf{0.591} \\
  & continuant  & 0.599 & 0.496 & 0.538& 0.451 \\
  & coronal  & \textbf{0.704} & \textbf{0.55} & \textbf{0.664} & \textbf{0.566} \\
   & distributed  & \textbf{0.744} & \textbf{0.555} &\textbf{0.68} &\textbf{0.587}  \\
   & sonorant  & 0.611 & 0.492 & 0.536& 0.447 \\
   & tap  & 0.684 & 0.548& 0.632&0.561 \\
   & voice  & 0.602 & 0.462  & 0.522& 0.422\\

\hline

\end{tabular}
}
\caption{Canonical correlation scores from SVCCA \cite{raghu2017svcca} between phonological feature probabilities and the subsets of relevant representations selected by the accent classifier probes; results are shown for the rhotic tap segment [\textfishhookr], with scores listed by accent, representation, and probe type (LR=Logistic Regression; SVM=Linear Support Vector Machine).}
\label{tab:ccatap}
\end{table}

\begin{table}[htpb]

\centering
\resizebox{\columnwidth}{!}{%
\begin{tabular}{cc|cc|cc}
\multicolumn{2}{c|}{} &\multicolumn{2}{c|}{\textbf{Wav2Vec2-BERT}} & \multicolumn{2}{c}{\textbf{WavLM}} \\
\hline
 \textbf{Accent} & \textbf{Feature} & \textbf{LR} &    \textbf{SVM} & \textbf{LR} &    \textbf{SVM}  \\

\hline
 No/Negligible &anterior & 0.511 & 0.649 &0.517 & 0.651   \\ 
   & consonantal  & \textbf{0.617} &  \textbf{0.771} &\textbf{0.643} & \textbf{0.787}\\
   & coronal  & 0.613 & 0.745& 0.617&  0.744\\ 
 \hline
   Mild &anterior & 0.655 & 0.737 & 0.624& 0.724  \\ 
   & consonantal  & \textbf{0.765} &  \textbf{0.827} & \textbf{0.754}& \textbf{0.839}\\
   & coronal  & 0.745 & 0.805 & 0.728& 0.799\\ 
\hline
 Strong &anterior & 0.661 & 0.731 & 0.609 & 0.732  \\ 
   & consonantal  & \textbf{0.776} &  \textbf{0.825} & \textbf{0.758}& \textbf{0.841}\\
   & coronal  & 0.746 & 0.799  & 0.715&0.806\\ 

\hline

\end{tabular}
}
\caption{Canonical correlation scores from SVCCA \cite{raghu2017svcca} between phonological feature probabilities and the subsets of relevant representations selected by the accent classifier probes; results are shown for the retroflex stop segment [\textrtailt], with scores listed by accent, representation, and probe type (LR=Logistic Regression; SVM=Linear Support Vector Machine).}
\label{tab:ccaretro}
\end{table}

\begin{table}[H]
\centering

\resizebox{\columnwidth}{!}{%
\begin{tabular}{cc|cc|cc}
\multicolumn{2}{c|}{} &\multicolumn{2}{c|}{\textbf{Wav2Vec2-BERT}} & \multicolumn{2}{c}{\textbf{WavLM}} \\
\hline
\textbf{Accent} & \textbf{Feature} & \textbf{LR} &    \textbf{SVM} & \textbf{LR} &    \textbf{SVM}  \\
\hline
 No/Negligible &  approximant & \textbf{0.573} & \textbf{0.756} &0.505 & \textbf{0.80} \\ 
   & consonantal  & 0.523 & 0.729 & \textbf{0.508} & 0.78 \\ 
   & continuant  & 0.494 & 0.73 & 0.481 & 0.776\\
   & delayed release  & 0.542 & 0.719 & 0.465 & 0.762 \\
   & labial  & 0.568 & 0.749 & 0.481 & 0.784\\
   & labiodental  & 0.565 & 0.741 & 0.477 & 0.78\\
   & sonorant  & \textbf{0.63} & \textbf{0.81} & \textbf{0.561} & \textbf{0.848}\\
   & voice  & \textbf{0.616} & \textbf{0.804} & \textbf{0.555} & \textbf{0.855}\\
 \hline
 Mild &  approximant & \textbf{0.718} &\textbf{0.836} &\textbf{0.672} & \textbf{0.882} \\ 
   & consonantal  & 0.676 & 0.812 & 0.653 & 0.856\\ 
   & continuant  & 0.683 & 0.813 &0.634 &0.864 \\
   & delayed release  & 0.684 & 0.799 &0.646 & 0.848\\
   & labial  & 0.699 & 0.823 & 0.66 & 0.871 \\
   & labiodental  & 0.70 & 0.819 & 0.667 & 0.862\\
   & sonorant  & \textbf{0.767} & \textbf{0.875} &\textbf{0.73} & \textbf{0.91}\\
   & voice  & \textbf{0.77} & \textbf{0.871} & \textbf{0.725} & \textbf{0.918}\\
 \hline
 Strong &  approximant & \textbf{0.739} & \textbf{0.842} &\textbf{0.699} &\textbf{0.878}  \\ 
   & consonantal  & 0.696 & 0.815 & 0.645& 0.851\\ 
   & continuant  & 0.701 & 0.815 &0.631 &0.864\\
   & delayed release  & 0.708 & 0.80 &0.677 & 0.846\\
   & labial  & 0.733 & 0.828 & 0.678& 0.866\\
   & labiodental  & 0.734 & 0.824 & 0.689 & 0.858\\
   & sonorant  & \textbf{0.784} & \textbf{0.872} &\textbf{0.74} & \textbf{0.911}\\
   & voice  & \textbf{0.779} &\textbf{0.867} & \textbf{0.727}&\textbf{0.916}\\
 \hline

\end{tabular}
}
\caption{Canonical correlation scores from SVCCA \cite{raghu2017svcca} between phonological feature probabilities and the subsets of relevant representations selected by the accent classifier probes; results are shown for the labiodental approximant segment [\textscriptv], with scores listed by accent, representation, and probe type (LR=Logistic Regression; SVM=Linear Support Vector Machine).}
\label{tab:ccalabiodental}
\end{table}

\subsection{Phone to phonological class mapping}\label{sec:appendix:phontoclassmapping}
Table ~\ref{tab:mapping} shows the merged mapping between the MFA phonesets from \citet{mfa_english_india_mfa_dictionary_2024,mfa_english_us_mfa_dictionary_2024} and the phonological features from \citet{hayes2011introductory}.

\begin{table*}[ht!]
\centering
\begin{tabular}{l|l}
\textbf{Phonological Feature} & \textbf{Phone List} \\
\hline
\hline
syllabic & a aj aw {a\textlengthmark} {e\textlengthmark} ej i {i\textlengthmark} {o\textlengthmark} ow {\ae} {\textturna} {\textscripta} {\textscripta\textlengthmark} {\textturnscripta} {\textturnscripta\textlengthmark} {\textopeno j} {\textschwa} {\textrhookschwa} {\textepsilon} {\textepsilon\textlengthmark} {\textrevepsilon} {\textrevepsilon\textlengthmark} {\textrhookrevepsilon} {\textsci} {\textbaru} {\textbaru\textlengthmark} {\textupsilon} \\
\hline
consonantal &  b {b\super j} c {c\super h} {c\super w} d {\textdyoghlig} {d\super j} \textsubbridge{d} f {f\super j} h j k {k\super h} {k\super w} l m {m\super j} \s{m} n \s{n} p {p\super h} {p\super j} {p\super w} s \\
& t {\textteshlig}  {t\super h} {t\super j}  {t\super w} \textsubbridge{t} v {v\super j} z \c{c} {\dh} {\textipa{\ng}} {\textrtaild} {\textbardotlessj} {\textbardotlessj\super w} g {g\super w} {\textbarl} {\s{\textbarl}} {\textltailm} {\textltailn} {\textfishhookr} {\textfishhookr\super j} {\~\textfishhookr} {\textesh} {\textrtailt} {\textrtailt\super j} {\textrtailt\super w} {\textturny} {\textyogh} {\textscriptv} {\textglotstop} {\texttheta} \\ 
\hline
long & {a\textlengthmark} {\textscripta\textlengthmark} {\textturnscripta\textlengthmark}
  {i\textlengthmark} {\textepsilon\textlengthmark} {{\textrevepsilon\textlengthmark}} {e\textlengthmark} {o\textlengthmark} {\textbaru\textlengthmark} \\
\hline
sonorant & a {a\textlengthmark} aj aw {\textturna} {\ae} {\textscripta} {\textscripta\textlengthmark}  {\textturnscripta} {\textturnscripta\textlengthmark} {\textepsilon} {\textepsilon\textlengthmark} {\textrevepsilon} {\textrevepsilon\textlengthmark} {\textrhookrevepsilon} {e\textlengthmark} ej   {\textsci} i {i\textlengthmark}  {o\textlengthmark} ow {\textopeno j} 
 {\textbaru} {\textbaru\textlengthmark} {{\textupsilon}} {\textschwa} {\textrhookschwa} \\
 & l {\textbarl} {\s{\textbarl}}  {\textturny} j {\textfishhookr} {\textfishhookr\super j} {\~\textfishhookr} {\textturnr} m {\s{m}} {m\super j}  {\textltailm}  {\textipa{\ng}} {\textltailn} {\s{n}} n {\textscriptv}  w \\

\hline
continuant & a {a\textlengthmark} aj aw {\textturna} {\ae} {\textscripta} {\textscripta\textlengthmark} {\textturnscripta} {\textturnscripta\textlengthmark}  {\textepsilon} {\textepsilon\textlengthmark} {\textrevepsilon} {\textrevepsilon\textlengthmark} {\textrhookrevepsilon} {e\textlengthmark} ej {\textsci} i {i\textlengthmark} {o\textlengthmark} ow {\textopeno j} {\textbaru} {\textbaru\textlengthmark} {\textupsilon}  {\textschwa} 
 {\textrhookschwa} \\ & 
 {\dh} {\texttheta} f {f\super j} j  {\textfishhookr} {\~\textfishhookr} {\textfishhookr\super j} {\textturnr}   {\textesh} {\textyogh} v {v\super j}  \c{c}  
l {\textbarl} {\s{\textbarl}} {\textturny} h s z   {\textscriptv} w \\
\hline
delayed release &  f {f\super j} {\textesh} {\textyogh} {\c{c}} v {v\super j} {\textteshlig} {\textdyoghlig}  h s z 
 {\dh} {\texttheta}\\
\hline
approximant & a {a\textlengthmark} aj aw {\textturna} {\ae} {\textscripta} {\textscripta\textlengthmark} {\textturnscripta} {\textturnscripta\textlengthmark}  {\textepsilon}  {\textepsilon\textlengthmark} {\textrevepsilon}  {\textrevepsilon\textlengthmark} {\textrhookrevepsilon} {e\textlengthmark} ej {\textsci} i {i\textlengthmark} {o\textlengthmark} ow {\textopeno j}  {\textbaru}  
 {\textbaru\textlengthmark} {\textupsilon} {\textschwa}  {\textrhookschwa}   \\
 & j  {\~{\textfishhookr}} 
 {\textfishhookr}  {\textfishhookr\super j} {\textturnr}      l {\s{\textbarl}} {\textbarl} {\textturny} {\textscriptv} w \\
\hline
tap & {\textfishhookr} {\~\textfishhookr} {\textfishhookr\super j} \\
\hline
nasal & m {m\super j} \s{m} {\textltailm} n \s{n} \textipa{\ng}  {\textltailn} \\
\hline
voice & a {a\textlengthmark} aj aw {\textturna} {\ae} {\textscripta} {\textscripta\textlengthmark} {\textturnscripta} {\textturnscripta\textlengthmark} {\textepsilon} {\textepsilon\textlengthmark} {\textrevepsilon } {\textrevepsilon\textlengthmark} {\textrhookrevepsilon} {e\textlengthmark} ej   {\textsci} i {i\textlengthmark} {o\textlengthmark} ow {\textopeno j} {\textbaru} {\textbaru\textlengthmark} {\textupsilon} {\textschwa} {\textrhookschwa} \\
& {\dh} d {d\super j} {\textrtaild} \textsubbridge{d} {\textfishhookr}  {\~\textfishhookr}  {\textfishhookr\super j}   {\textturnr}  
 j {\textbardotlessj} {\textbardotlessj\super w}  {\textyogh} {\textdyoghlig}  v {v\super j} 
m {m\super j} {\s{m}} {\textltailm} n {\textipa{\ng}} {\s{n}} {\textltailn}  b {b\super j} l {\s{\textbarl}}  {\textturny} {g\super w} g z {\textscriptv} w \\
\hline
spread glottis & h \\
\hline
labial & p {p\super j} {p\super h} {p\super w} f {f\super j}    v {v\super j}  {\textscriptv}  {\textltailm} {m\super j} m {\s{m}} b {b\super j} \\
\hline
round &  {\textturnscripta} {\textturnscripta\textlengthmark}  ow  {o\textlengthmark} {\textopeno j} {\textbaru} {\textbaru\textlengthmark} {\textupsilon}  \\
\hline
labiodental & f {f\super j} {\textltailm} v {v\super j} {\textscriptv} \\
\hline
coronal &  c {c\super h}  {c\super w}  {\c{c}}   {\textfishhookr} {\~\textfishhookr} {\textfishhookr\super j} {\textturnr}  {\textesh} {\textyogh} {\textdyoghlig}   {\textteshlig} {\textrtailt} {\textrtailt\super j} {\textrtailt\super w}  t {t\super h} {t\super w} {t\super j}  \textsubbridge{t} {\s{n}} n {\textltailn} {\textturny}  d {\textrtaild} \textsubbridge{d} {d\super j}  l {\textbarl} {\s{\textbarl}} s z {\texttheta}  {\dh}\\
\hline
anterior &   {\textfishhookr} {\~\textfishhookr} {\textfishhookr\super j} t {t\super w} {t\super h} {t\super j}  {\textsubbridge{t}} d {d\super j} {\textsubbridge{d}} {\s{n}} n l {\textbarl} {\s{\textbarl}} s z  {\texttheta} {\dh} \\
\hline
distributed &  c {\c{c}} {c\super h} {c\super w}{\textbardotlessj} {\textbardotlessj\super w}  {\textteshlig} {\textdyoghlig}   {\textesh} {\textyogh} {\textturnr} {\textturny}  {\textltailn} {\texttheta} {\dh} \\
\hline
strident & s z {\textteshlig} {\textdyoghlig} {\textesh}  {\textyogh} \\ 
\hline
lateral & l {\textbarl} {\s{\textbarl}} {\textturny} \\
\hline
dorsal & a {a\textlengthmark} aj aw {\textturna} {\ae} {\textscripta} {\textscripta\textlengthmark} {\textturnscripta} {\textturnscripta\textlengthmark} 
 {\textepsilon} {\textepsilon\textlengthmark}  {\textrevepsilon} {\textrevepsilon\textlengthmark} {\textrhookrevepsilon} {e\textlengthmark} ej {\textsci} i {i\textlengthmark} {o\textlengthmark} ow {\textopeno j} 
  {\textbaru} {\textbaru\textlengthmark} {\textupsilon} {\textschwa} {\textrhookschwa} \\
  & c {c\super h} {c\super w} {\c{c}} k {k\super w} g {g\super w}  {\textipa{\ng}}  
    {\textltailn}   {\textbarl} {\s{\textbarl}} {\textturny} w \\
\hline
high & {\textsci} i {i\textlengthmark} {\textbaru}  {\textbaru\textlengthmark} {\textupsilon}  c {c\super h} {c\super w} {\c{c}} k {k\super w} g {g\super w} {\textturny} {\textipa{\ng}} {\textltailn} w \\
\hline
low & a {a\textlengthmark} aj aw {\textscripta} {\textscripta\textlengthmark} {\textturnscripta} {\textturnscripta\textlengthmark} {\ae}     \\
\hline
front & {\ae} {\textepsilon} {\textepsilon\textlengthmark} \textsci  i {i\textlengthmark} c {c\super h} {c\super w} {\c{c}} {e\textlengthmark} ej  j {\textbardotlessj} {\textbardotlessj\super w} {\textltailn} {\textturny} \\
\hline
back & {\textscripta} {\textscripta\textlengthmark} {\textturnscripta} {\textturnscripta\textlengthmark} {\textrevepsilon} {\textrevepsilon\textlengthmark} {\textrhookrevepsilon} {o\textlengthmark} ow {\textopeno j} {\textupsilon} {\textbarl} {\s{\textbarl}} w\\
\hline
tense & {e\textlengthmark} ej i {i\textlengthmark} {\textbaru} {\textbaru\textlengthmark} {o\textlengthmark} ow {\textschwa} {\textrhookschwa} j w\\
\hline
constr. glottis & {\textglotstop}\\

\end{tabular}
\caption{Mapping between MFA phonesets and phonological features from \citet{hayes2011introductory} for Phonet modeling.}
\label{tab:mapping}
\end{table*}

\end{document}